\documentclass[superscriptaddress,amsmath, nofootinbib]{revtex4}
\usepackage{amsfonts}
\usepackage{graphicx}
\usepackage[latin1]{inputenc}
\usepackage{amsmath}
\usepackage{amssymb}
\begin{document}


\title{Rotation matrix of a charged symmetrical body: one-parameter family of solutions in elementary functions.}

\author{Alexei A. Deriglazov }
\email{alexei.deriglazov@ufjf.br} \affiliation{Depto. de Matem\'atica, ICE, Universidade Federal de Juiz de Fora,
MG, Brazil} 

\author{}

\date{\today}

\begin{abstract}
Euler-Poisson equations of a charged spinning body in external constant and homogeneous electric and magnetic fields are  deduced starting from the variational problem, where the body is considered as a system of charged point particles subject to holonomic constraints. The final equations are written for the  center-of mass-coordinate, rotation matrix and angular velocity.  General solution to the equations of motion is obtained for the case of a charged ball. For the case of a symmetrical charged body (solenoid), the task of obtaining the general solution to Euler equations is reduced to the problem of a one-dimensional cubic pseudo-oscillator. Besides, we present a one-parametric family of solutions to the problem in elementary functions. 
\end{abstract}

\maketitle 



\section{Introduction.}\label{Intr}

Spinning bodies represent an important object of study in modern research at different scales, from problems of approaching compact stars in gravitational wave physics to levitating nano-particles and elementary particles with spin. Behavior of a spinning object in special and general relativity as well as in quantum mechanics in many cases is considered in first order by perturbation theory based on exact solutions to the classical problem \cite{Ham_22}. Therefore, the search for new integrable cases and analytical solutions in the dynamics of a spinning body is important for further progress in such problems \cite{Huss_2019, Tikn_2009, Mor_2023}. In the present work we consider a charged spinning body in external constant and homogeneous electric and magnetic fields. While the case of a ferromagnet has been discussed quite widely in the literature \cite{Mar_1980, Mar_1981, Sam_1984, Koz_1985, Zho_2009, Ame_2016}, much less attention has been paid to a charged dielectric \cite{Gri_1957}. In this article we will try to fill this gap in the literature.

The work is organized as follows. In Sect. \ref{CH6.3} we deduce equations of motion on the base of a Lagrangian action, formulated for the case under consideration. 
We will present a detailed derivation of the equations, since semi-empirical methods applied to a spinning body in some cases lead to either inaccuracies or erroneous interpretation of the final result \cite{AAD23, AAD23_3, AAD23_8, AAD24_1}. In Sect. \ref{CH6.4} we present general solution to obtained equations for the case of a charged ball. In Sect. \ref{CH6.5} we reduce the problem of a symmetrical charged body to the problem of a one-dimensional non-linear pseudo-oscillator. In Sect. \ref{CH6.6} we present a one-parametric family of solutions in elementary functions for the motions with specially chosen initial angular velocity of the symmetrical charged body. In the Appendix, for the convenience of the reader, we summarized the motion of a point charged particle in external electromagnetic field.

\section{Charged body in constant and homogeneous electric and magnetic fields.}\label{CH6.3}

Consider a rigid body that consist of $n$ particles ${\bf y}_N(t)$ of charge $e_N$ and mass $m_N$, $N=1, 2, \ldots , n$. Its Lagrangian action reads \cite{AAD23}
\begin{eqnarray}\label{1.6}
S=\int dt ~\frac12\sum_{N=1}^{n}m_N\dot{\bf y}_N^2+\frac12\sum_{A, B=2}^{4}\lambda_{AB}\left[({\bf y}_A-{\bf y}_1, {\bf y}_B-{\bf y}_1)-a_{AB}\right]
+\frac12\sum_{A=2}^{4}\sum_{\beta=5}^{n}\lambda_{A\beta}\left[({\bf y}_A-{\bf y}_1, {\bf y}_\beta-{\bf y}_1)-a_{A\beta}\right]. 
\end{eqnarray}
The first term is kinetic energy of all particles, while the remaining  terms account the presence of constraints, that guarantee that distances and angles among the particles do not change with time\footnote{We use the notation from \cite{AAD23}. In particular, by $({\bf a}, {\bf b})$ and $[{\bf a}, {\bf b}]_i=\epsilon_{ijk}a_j b_k$ we denote the scalar and vector products of the vectors ${\bf a}$ and ${\bf b}$. $\epsilon_{ijk}$ is Levi-Civita symbol in three dimensions, with $\epsilon_{123}=1$.}.  The constraints were added with help of Lagrangian multipliers  $\lambda_{AN}(t)$.  In all calculations these auxiliary variables should be treated on equal footing with ${\bf y}_N(t)$. In particular, looking for the equations of motion, we take variations with respect to ${\bf y}_N$ and  all $\lambda_{AN}$. The $3\times3$\,-block $\lambda_{AB}$ of $\lambda_{AN}$ was chosen to be the symmetric matrix. The variations with respect  to $\lambda_{AN}$ imply the constraints, which therefore arise as a part of conditions of extreme of the action functional. So the presence  of $\lambda_{AN}$ allows $y_N^i$ to be treated as unconstrained variables, that should be varied independently in obtaining the equations of motion. 

We consider the body immersed into constant and homogeneous electric and magnetic fields with scalar potential $A_0({\bf y}_N)$ and vector potential ${\bf A}({\bf y}_N)$, see Appendix.  Generally, its movement will produce time-dependent distribution of charges and currents leading to radiation of an electromagnetic fields, see Chapter 17 in \cite{Jac_1975}. We neglect the resulting radiation and radiation damping effects. Then,  summing up the potential energies (\ref{mf13}) of body's particles, we obtain its total potential energy 
\begin{eqnarray}\label{mf15}
-U=\sum_{N=1}^n e_NA_0({\bf y}_N)+\frac{e_N}{c}({\bf A}({\bf y}_N), \dot{\bf y}_N). 
\end{eqnarray}
Adding it to the action (\ref{1.6}), we obtain a variational problem for the body in external electric and magnetic fields. We assume that all particles of the body have the same charge to mass ratio, $\frac{e_N}{m_N}=\mu$ for any $N=1, 2, \ldots , n$. Then our action implies the following dynamical equations
\begin{eqnarray}\label{mf16}
m_1\ddot{\bf y}_1=-\sum_{AB}\lambda_{AB}[{\bf y}_B-{\bf y}_1]-\frac12\sum_{A\alpha}\lambda_{A\alpha}[{\bf y}_A+{\bf y}_\alpha-2{\bf y}_1]+
\mu m_1{\bf E}+\frac{\mu}{c}m_1[\dot{\bf y}_1, {\bf B}], \cr
m_A\ddot{\bf y}_A=\sum_{B}\lambda_{AB}[{\bf y}_B-{\bf y}_1]+\frac12\sum_{\alpha}\lambda_{A\alpha}[{\bf y}_\alpha-{\bf y}_1]+\mu m_A{\bf E}+\frac{\mu}{c}m_A[\dot{\bf y}_A, {\bf B}], \cr
m_\alpha\ddot{\bf y}_\alpha=\frac12\sum_{A}\lambda_{A\alpha}[{\bf y}_A-{\bf y}_1]+\mu m_\alpha{\bf E}+
\frac{\mu}{c}m_\alpha[\dot{\bf y}_\alpha, {\bf B}].
\end{eqnarray}
Introducing the center of mass, ${\bf y}_0=\frac{1}{\mu_0}\sum m_N{\bf y}_N$, 
where $\mu_0=\sum m_N$, the equations (\ref{mf16}) imply
\begin{eqnarray}\label{mf17}
\ddot{\bf y}_0=\mu{\bf E}+\frac{\mu}{c}[\dot{\bf y}_0, {\bf B}],
\end{eqnarray}
that is center-of-mass behaves like charged point particle discussed in Appendix. In particular, rotational motion of the body does not affect its translational motion. 

Substituting ${\bf y}_N={\bf y}_0+{\bf x}_N$ into Eq. (\ref{mf16}) and taking into account (\ref{mf17}),  we rewrite these equations in the center-of-mass coordinate system 
\begin{eqnarray}\label{mf18}
m_1\ddot x_1^i=-\sum_{A,B=2}^4\lambda_{AB}[x_B^i-x_1^i]-\frac12\sum_{A\alpha}\lambda_{A\alpha}[x_A^i+x_\alpha^i-2x_1^i]+m_1\lambda^i+
\mu m_1{\bf E}+\frac{\mu}{c}m_1[\dot{\bf x}_1, {\bf B}], \cr
m_A\ddot x_A^i=\sum_{B=2}^4\lambda_{AB}[x_B^i-x^i_1]+\frac12\sum_{\alpha=5}^n\lambda_{A\alpha}[x_\alpha^i-x_1^i]+m_A\lambda^i+
\mu m_A{\bf E}+\frac{\mu}{c}m_A[\dot{\bf x}_A, {\bf B}], \cr
m_\alpha\ddot x_\alpha^i=\frac12\sum_{A=2}^4\lambda_{A\alpha}[x_A^i-x_1^i]+m_\alpha\lambda^i+
\mu m_\alpha{\bf E}+\frac{\mu}{c}m_\alpha[\dot{\bf x}_\alpha, {\bf B}].
\end{eqnarray}
Each solution $x^i_N(t)$ to these equations is of the form $x^i_N(t)=R_{ik}(t)x^k_N(0)$, where $R_{ij}(t)$ is an orthogonal matrix that, by construction, obeys the universal initial data $R_{ij}(0)=\delta_{ij}$. Substituting this expression into Eqs. (\ref{mf18}), then multiplying the equation with number $N$ by $x_N^j(0)$  and taking their sum, we obtain the following second-order equations for determining the rotation matrix $R_{ik}$: 
\begin{eqnarray}\label{mf19}
\ddot R_{ik}g_{kj}=-R_{ik}\lambda_{kj}+\mu E_i(\sum m_N{\bf x}_N)-A_{ia}g_{aj}, \qquad R^TR={\bf 1}, \qquad R(0)={\bf 1}.
\end{eqnarray}
It was denoted
\begin{eqnarray}\label{mf20}
A_{ia}=\frac{\mu}{c}\epsilon_{inp}B_n\dot R_{pa}.
\end{eqnarray}
Besides, $g_{ik}$ is the mass matrix 
\begin{eqnarray}\label{1.32}
g_{ij}\equiv\sum_{N=1}^{n}m_Nx_N^i(0)x_N^j(0),
\end{eqnarray}
while $\lambda_{jk}(t)$ is the symmetric matrix 
\begin{eqnarray}\label{1.38}
\lambda_{jk}=-\sum_{AB}\lambda_{AB}\left[x_1^jx_1^k+x_A^jx_B^k-x_B^{(j}x_1^{k)}\right]-\frac12\sum_{A\alpha}\lambda_{A\alpha}\left[x_\alpha^{(j}x_A^{k)}-x_\alpha^{(j}x_1^{k)}-x_A^{(j}x_1^{k)}-2x_1^{j}x_1^{k}\right],
\end{eqnarray}
where all $x_N^i$ are taken at the instant $t=0$. 

Due to the  identity $\sum m_N{\bf x}_N=0$, satisfied for the center-of-mass coordinates, second term on r. h. s. of Eq. (\ref{mf19}) vanishes. In the result,  the electric field do not affect the motion of rotational degrees of freedom. Besides, the center-of-mass variable do not enter into this equation, so the translational motion does not affect the rotational motion of the body. 

The variables $\lambda_{jk}(t)$ in equations (\ref{mf19}) depend on the unknown dynamical variables $\lambda_{AN}(t)$. Fortunately,  we do not need to 
know $\lambda_{jk}(\lambda_{AN}(t))$, because of these equations determine $\lambda_{jk}$ algebraically, as some functions of $R$ and $\dot R$.  This result, obtained with use of the procedure described in \cite{AAD23}, can be formulated as follows \par 

\noindent {\bf Affirmation.}\label{Aff100} Consider the second-order system for determining the variables $R_{ij}(t)$ and $\lambda_{kj}(t)$ 
\begin{eqnarray}\label{The.1}
\ddot R_{ik}g_{kj}=-R_{ik}\lambda_{kj}-A_{ik}g_{kj}, \qquad R^TR={\bf 1}, 
\end{eqnarray}
where $A_{ik}(R_{ab}, \dot R_{ab})$ is some given matrix that does not depend on $g_{kj}$ and $\lambda_{kj}$ (as before, $g_{kj}$ is a numerical symmetric non-degenerate matrix, and $\lambda_{kj}(t)=\lambda_{jk}(t)$). 

The problem (\ref{The.1}) is equivalent to the following Cauchy problem for the first-order system, written for the mutually independent variables $R_{ij}(t)$ and $\Omega_i(t)$: 
\begin{eqnarray}\label{mf22}
I_{ka}\dot\Omega_a=[I{\boldsymbol\Omega}, {\boldsymbol\Omega}]_k+\frac12\epsilon_{kij}\left[(AI)_{ai}R_{aj}+(RI)_{ai}A_{aj}\right]- 
\frac12 I_{ka}\epsilon_{aij}A_{bi}R_{bj}, 
\end{eqnarray}
\begin{eqnarray}\label{mf21}
\dot R_{ij}=-\epsilon_{jab}\Omega_a R_{ib}, \qquad R_{ij}(0)=\delta_{ij}, 
\end{eqnarray}
where $I$ is inertia tensor of the body with the components $I_{ij}=(\sum g_{aa})\delta_{ij}-g_{ij}$. 

Note that this system is composed of $SO(3)$ vectors and tensors, so it is covariant under the rotations.  We will work with these equations assuming that the mass matrix and inertia tensor are of diagonal form. This implies \cite{AAD23}, that at initial instant $t=0$ the Laboratory basis vectors ${\bf e}_i$ were taken in the directions of axes of inertia ${\bf R}_i(t)$ taken as the body-fixed frame: ${\bf e}_i={\bf R}_i(0)$. We also recall that the body-fixed basis vectors are columns of the rotation matrix: $R=({\bf R}_1,  {\bf R}_2, {\bf R}_3)$. Eigenvalues of mass matrix and inertia tensor are related as follows: $2g_1=I_2+I_3-I_1$, and so on.  

Our equations (\ref{mf19}) are of the form (\ref{The.1}), so they are equivalent to  the first-order system (\ref{mf22}) and (\ref{mf21}) with $A$ written in (\ref{mf20}).  To obtain an explicit form of Eq. (\ref{mf22}), we use (\ref{mf21})  to rewrite the quantity $A_{ia}$ as follows:
\begin{eqnarray}\label{mf23}
\epsilon_{ijk}B_j\dot R_{ka}=\epsilon_{ijk}\epsilon_{abc}B_j\Omega_b R_{kc}=\epsilon_{ijk}\epsilon_{\alpha\beta\gamma}R_{\alpha a}R_{\beta b}R_{\gamma c}B_j\Omega_b R_{kc}= \qquad \qquad \qquad \cr \epsilon_{ijk}\epsilon_{\alpha\beta\gamma}R_{\alpha a}B_j(R{\boldsymbol\Omega})_\beta (RR^T)_{\gamma k}= 
\epsilon_{ijk}\epsilon_{\alpha\beta k}R_{\alpha a}B_j(R{\boldsymbol\Omega})_\beta=R_{ia}({\bf B}R{\boldsymbol\Omega})-(R{\boldsymbol\Omega})_i({\bf B}R)_a. \nonumber
\end{eqnarray}
Thus
\begin{eqnarray}\label{mf24}
A_{ia}=-\frac{\mu}{c}\epsilon_{ijk}\epsilon_{abc}B_j\Omega_b R_{kc}=\frac{\mu}{c}\left[(R{\boldsymbol\Omega})_i({\bf B}R)_a-R_{ia}({\bf B}R{\boldsymbol\Omega})\right].
\end{eqnarray}
Using the latter expression for $A_{ia}$ in Eq. (\ref{mf22}), after direct calculations this acquires the form
\begin{eqnarray}\label{mf25}
I\dot{\boldsymbol\Omega}=[I{\boldsymbol\Omega}, {\boldsymbol\Omega}]-\frac{\mu}{2c}\left\{I[R^T{\bf B}, {\boldsymbol\Omega}]-[IR^T{\bf B}, {\boldsymbol\Omega}]+[R^T{\bf B}, I{\boldsymbol\Omega}]\right\}. 
\end{eqnarray}
The vector composed of last three terms can be written in a more compact form in terms of the mass matrix. Indeed, writing the first component of this vector in explicit form  we get 
\begin{eqnarray}\label{mf26}
(I[R^T{\bf B}, {\boldsymbol\Omega}])_1-[IR^T{\bf B}, {\boldsymbol\Omega}]_1+[R^T{\bf B}, I{\boldsymbol\Omega}]_1= 
(I_1-I_2+I_3)(R^T{\bf B})_2\Omega_3-(I_1-I_3+I_2)(R^T{\bf B})_3\Omega_2= \cr 2g_2(R^T{\bf B})_2\Omega_3-2g_3(R^T{\bf B})_3\Omega_2=
2[gR^T{\bf B}, {\boldsymbol\Omega}]_1, \qquad \qquad \qquad
\end{eqnarray}
and similar expressions for the second and third components. Then Eq. (\ref{mf22}) acquires the final form
\begin{eqnarray}\label{mf27}
I\dot{\boldsymbol\Omega}=[I{\boldsymbol\Omega}, {\boldsymbol\Omega}]-\frac{\mu}{c}[gR^T{\bf B}, {\boldsymbol\Omega}].
\end{eqnarray}
For completeness we also present equations for the vector of angular momentum ${\bf m}=RI{\boldsymbol\Omega}$ and for its 
components ${\bf M}=I{\boldsymbol\Omega}$  in the body-fixed 
frame 
\begin{eqnarray}
\dot{\bf m}=-\frac{\mu}{c}[RgR^T{\bf B}, RI^{-1}R^T{\bf m}], \qquad ~ \label{mf28} \\
\dot{\bf M}=[{\bf M}, I^{-1}{\bf M}]-\frac{\mu}{c}[gR^T{\bf B}, I^{-1}{\bf M}]. \label{mf29}
\end{eqnarray}

Let us consider Eqs. (\ref{mf27}) in the Laboratory system with third axis in the direction of magnetic vector ${\bf B}$. Denote by  $I'$ and $g'$ the non diagonal tensors of inertia and mass that will appear in this coordinates. Then Eqs. (\ref{mf27}) acquire the 
form $I'\dot{\boldsymbol\Omega}=[I'{\boldsymbol\Omega}, {\boldsymbol\Omega}]-\frac{\mu|{\bf B}|}{c}[g'{\bf G}_3, {\boldsymbol\Omega}]$, where ${\bf G}_3$ is the third row of the rotation matrix $R_{ij}$. They coincide with those deduced by G. Grioli, see \cite{Gri_1957, Ham_22}. 

In resume, we have succeeded in obtaining equations of motion (\ref{mf27}), (\ref{mf21}) and (\ref{mf17}) for a spinning charged body in external electric and magnetic fields. The solution $R_{ij}(t)$, $y_0^i(t)$ to these equations contains complete information on the evolution of the body with respect to Laboratory frame: dynamics of the body's point $y_N^i(t)$ with initial position $y_{N}^i(0)=y_0^i(0)+x_N^i(0)$ is $y_N^i(t)=y_0^i(t)+R_{ij}(t)x_{N}^j(0)$.

\section{General solution to the equations of a charged ball.}\label{CH6.4} 
Consider a totally symmetric charged body: 
\begin{eqnarray}\label{mf30.0}
I_1=I_2=I_3\equiv I_0. 
\end{eqnarray}
This could be a charged ball. Then its center moves 
according to Eq.  (\ref{mf10.2}). From kinematic relations ${\bf m}=\sum_{N=1}^{n}m_N[{\bf x}_N, \dot{\bf x}_N]=\sum_i g_i[{\bf R}_i, \dot{\bf R}_i]=RIR^T{\boldsymbol\omega}=RI{\bf\Omega}$ between angular momentum ${\bf m}$, angular velocity ${\boldsymbol\omega}$ and its 
components $\Omega_i$ in the body-fixed frame, together with Eq. (\ref{mf30.0}), we get ${\bf m}=I_0{\boldsymbol\omega}=I_0 R{\boldsymbol\Omega}$. The first equality means that for all $t$ the instantaneous rotation axis ${\boldsymbol\omega}$ remains parallel with the vector of angular momentum. Substituting (\ref{mf30.0}) into equations of previous section, we get 
\begin{eqnarray}
\dot R_{ij}=-\epsilon_{jab}\Omega_a R_{ib}, \qquad R_{ij}(0)=\delta_{ij},  \label{mf30} \\
\dot{\boldsymbol\Omega}=-\frac{\mu}{2c}[R^T{\bf B}, {\boldsymbol\Omega}]; \qquad \qquad \qquad \quad ~ \label{mf31}
\end{eqnarray}
\begin{eqnarray}\label{mf32}
\dot{\bf m}=-\frac{\mu}{2c}[{\bf B}, {\bf m}].  
\end{eqnarray}
The last equation implies that angular momentum precesses around ${\bf B}$ with Larmor's frequency $\alpha=\frac{\mu|{\bf B}|}{2c}$. 
Besides, the length of angular momentum and its projection on ${\bf B}$\,-axis are first integrals 
\begin{eqnarray}\label{mf33}
{\bf m}^2=\mbox{const}, \qquad ({\bf m}, {\bf B})=\mbox{const}. 
\end{eqnarray}
Components $\Omega_i$ of angular velocity ${\boldsymbol\omega}$ with respect to body-fixed frame precess with the same frequency around the 
components $(R^T{\bf B})_i$ of magnetic field ${\bf B}$ in body-fixed frame. 

Substituting ${\boldsymbol\Omega}=\frac{1}{I_0}R^T{\bf m}$ into (\ref{mf31}) and taking into account (\ref{mf30}), we arrive at the equation (\ref{mf32}). So the system (\ref{mf30}), (\ref{mf31}) is equivalent 
to $\dot R_{ij}=-\frac{1}{I_0}\epsilon_{jab}(R^T{\bf m})_a R_{ib}$, $\dot{\bf m}=-\frac{\mu}{2c}[{\bf B}, {\bf m}]$. Further, the first (non linear on $R$) equation of latter system can be replaced on the linear equation $\dot R_{ij}=\frac{1}{I_0}\epsilon_{iab}m_a R_{bj}$. The resulting system an the initial one have the same solutions in the set of orthogonal matrices. In the result, instead of Eqs. (\ref{mf30}) and (\ref{mf31}), a totally symmetric body can be described by the equations 
\begin{eqnarray}
\dot R_{ij}=\frac{1}{I_0}\epsilon_{iab}m_a R_{bj}, \qquad \mbox{or, equivalently} \quad \dot{\bf R}_j=\frac{1}{I_0}[{\bf m}, {\bf R}_j],  \label{mf34} \\
\dot{\bf m}=-\frac{\mu}{2c}[{\bf B}, {\bf m}],\qquad \qquad \qquad \qquad  \qquad \qquad \qquad \qquad  \qquad \label{mf35}
\end{eqnarray}
where all quantities are defined with respect to the Laboratory system. 
The initial conditions are $R_{ij}(0)=\delta_{ij}$. According to these equations, ${\bf m}(t)$ precesses around constant vector ${\bf B}$, while the vectors ${\bf R}_j(t)$ instantaneously precess around ${\bf m}(t)$. 

\begin{figure}[t] \centering
\includegraphics[width=06cm]{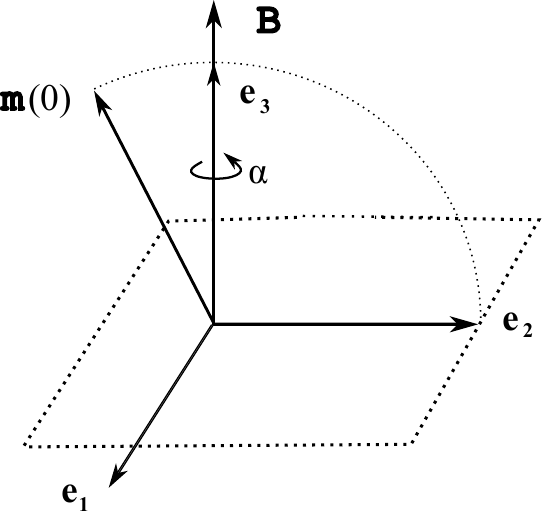}
\caption{Choice of Laboratory system ${\bf e}_i$ for analysis of a charged ball subject to electric and magnetic fields.}\label{Ball}
\end{figure}

Let us obtain the general solution to the system (\ref{mf34}), (\ref{mf35}). For a totally symmetric body we can choose the directions of Laboratory axes as convenient, this will not violate the diagonal form of inertia tensor. Using this freedom, we choose the Laboratory system so that at $t=0$ the vectors ${\bf B}$ and ${\bf m}(0)$ lie in the plane of ${\bf e}_2$, and ${\bf e}_3$, and ${\bf B}$ is directed along ${\bf e}_3$, see Figure \ref{Ball}. 
Then ${\bf B}=(0, 0, |{\bf B}|)^T$, ${\bf m}(0)=(0, m_2, m_3)^T$, and  
\begin{eqnarray}\label{mf36}
{\bf m}(t)=\left(
\begin{array}{ccc}
\cos\alpha t & \sin\alpha t & 0 \\
-\sin\alpha t & \cos\alpha t & 0 \\
0 & 0 & 1
\end{array}\right)\left(
\begin{array}{c}
0 \\
m_2 \\
m_3
\end{array}\right)=\left(
\begin{array}{c}
m_2\sin\alpha t \\
m_2\cos\alpha t \\
m_3
\end{array}\right),
\end{eqnarray}
is  a solution to Eq. (\ref{mf35}), where the precession frequency is the Larmor's frequency 
\begin{eqnarray}\label{mf37}
\alpha=\frac{\mu|{\bf B}|}{2c}.
\end{eqnarray}
To solve Eq. (\ref{mf34}) with this ${\bf m}(t)$, we look for its solution in the form
\begin{eqnarray}
\left(
\begin{array}{ccc}
\cos \gamma t\cos\alpha t-\hat k_3\sin \gamma t\sin\alpha  t & \hat k_3 \sin \gamma t\cos\alpha  t+(\hat k_2^2 +\hat k_3^2\cos \gamma t)\sin\alpha  t & -\hat k_2 \sin \gamma t\cos\alpha  t+\hat k_2 \hat k_3(1-\cos \gamma t)\sin\alpha  t \\
{} & {} & {} \\
-\cos \gamma t\sin\alpha  t-\hat k_3\sin \gamma t\cos\alpha  t  &   -\hat k_3 \sin \gamma t\sin\alpha  t+(\hat k_2^2 +\hat k_3^2\cos \gamma t)\cos\alpha  t & \hat k_2 \sin \gamma t\sin\alpha  t+\hat k_2 \hat k_3(1-\cos \gamma t)\cos\alpha  t \\
{} & {} & {} \\
\hat k_2\sin \gamma t & \hat k_2 \hat k_3(1-\cos \gamma t) & \hat k_3^2 +\hat k_2^2\cos \gamma t 
\end{array}\right)\equiv  \nonumber   {} & {} & {} \\ \nonumber 
\end{eqnarray}
\begin{eqnarray}\label{mf38}
\left(
\begin{array}{ccc} 
\cos \alpha t  & \sin \alpha t & 0  \\ 
-\sin \alpha t &  \cos \alpha t & 0 \\
0 & 0 & 1 \\
\end{array}\right)\times\left(
\begin{array}{ccc}
\cos \gamma t & \hat k_3\sin \gamma t  &  -\hat k_2\sin \gamma t  \\
{} & {} & {} \\
-\hat k_3 \sin \gamma t &  \hat k_2^2 +\hat k_3^2\cos \gamma t & \hat k_2\hat k_3(1-\cos \gamma t) \\
{} & {} & {} \\
\hat k_2 \sin \gamma t & \hat k_2 \hat k_3(1-\cos \gamma t) & \hat k_3^2 +\hat k_2^2\cos \gamma t
\end{array}\right). \qquad \qquad
\end{eqnarray}
To fix $\hat{\bf k}$ and $\gamma$, we substitute the ansatz (\ref{mf38}) into (\ref{mf34}) an then take $t=0$ in the resulting expressions. They determine $\hat{\bf k}$ as follows: $\hat k_2=-\frac{m_2}{\gamma I_0}$, $\hat k_3=-\frac{\alpha I_0+m_3}{\gamma I_0}$. Then $\hat k_2^2+\hat k_3^2=1$ implies $I_0^2\gamma^2=m_2^2+(\alpha I_0+m_3)^2$. The obtained equalities allow us to represent ${\bf k}$ and $\gamma$ 
through ${\bf m}$ and $\alpha$ as follows:
\begin{eqnarray}\label{mf39}
\hat k_2=-\frac{m_2}{\sqrt{m_2^2+(\alpha I_0+m_3)^2}}, \qquad \hat k_3=-\frac{m_3+\alpha I_0}{\sqrt{m_2^2+(\alpha I_0+m_3)^2}}, \qquad 
\gamma=\frac{1}{I_0}\sqrt{m_2^2+(\alpha I_0+m_3)^2}. 
\end{eqnarray}
By direct calculations, it can be verified that the expression (\ref{mf38}) with these $\hat{\bf k}$ and $\gamma$ satisfies the equations (\ref{mf34}).

In resume, we obtained analytical solution for a charged ball launched with initial angular velocity ${\boldsymbol\omega}=\frac{1}{I_0}{\bf m}=\frac{1}{I_0}(0, m_2, m_3)^T$ in constant and homogeneous electric ${\bf E}$ and magnetic ${\bf B}$ fields.    
It is given by the double-frequency rotation matrix (\ref{mf38}), (\ref{mf39}). The total motion can be thought as a superposition of two rotations: the first around unit vector $\hat{\bf k}=(0, \hat k_2, \hat k_3)^T$ with the frequency $\gamma$, and the second around the axis of magnetic 
field ${\bf B}=(0, 0, |{\bf B}|)^T$ with the frequency $\alpha=\frac{\mu|{\bf B}|}{2c}$. The angular momentum vector ${\bf m}(t)$ precesses around the vector ${\bf B}$ with the Larmor's frequency $\alpha=\frac{\mu|{\bf B}|}{2c}$. 

Let us consider the ball launched with initial vector of angular velocity parallel to the vector of magnetic field ${\bf B}$. That is the initial conditions are ${\bf m}(0)=(0, 0, m_3)^T$. Then ${\bf m}(t)=(0, 0, m_3)^T$, $\hat{\bf k}=(0, 0, -1)^T$ 
and $\gamma=\alpha+\frac{m_3}{I_0}$. With these values, the rotation matrix (\ref{mf38}) reduces to 
\begin{eqnarray}\label{mf40}
R=\left(
\begin{array}{ccc}
\cos \frac{m_3}{I_0} t  & -\sin \frac{m_3}{I_0} t & 0  \\ 
\sin \frac{m_3}{I_0} t &  \cos \frac{m_3}{I_0} t & 0 \\
0 & 0 & 1 \\
\end{array}\right).
\end{eqnarray}
As it should be expected, the ball experiences a stationary rotation around the vector of magnetic field ${\bf B}$ with the frequency $\frac{m_3}{I_0}$.

\section{Symmetrical charged body and one-dimensional non linear pseudo-oscillator.}\label{CH6.5}  

In this section we start to study the symmetrical charged body. We show that for any solution $R_{ij}(t)$, $\Omega_i(t)$ to Euler-Poisson equations, the function $\Omega_3(t)$ obeys to the equation of a one-dimensional cubic pseudo-oscillator, see Eq. (\ref{mf55}) below. Besides, when $\Omega_3(t)$ is known, the functions $\Omega_1(t)$ and $\Omega_2(t)$ can be found by quadratures, see Eqs. (\ref{mf52}) and (\ref{mf53}). 

\begin{figure}[t] \centering
\includegraphics[width=08cm]{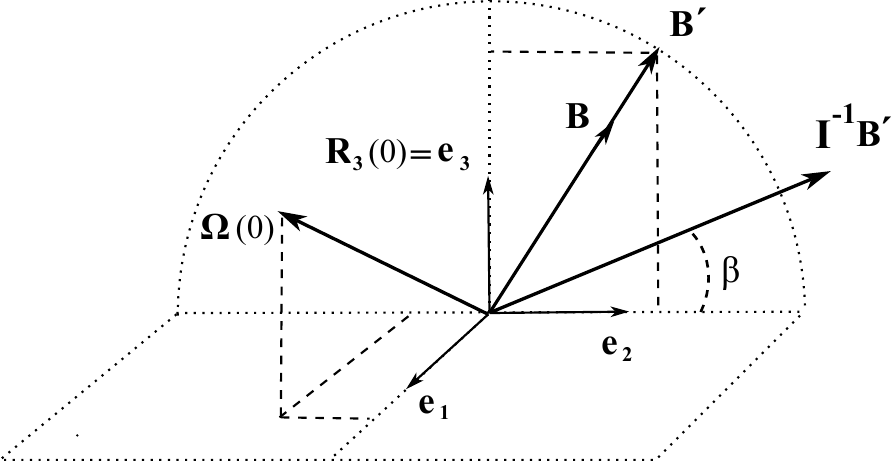}
\caption{Choice of Laboratory system for analysis of a charged symmetrical body.}\label{Lssc}
\end{figure}

Consider Eqs. (\ref{mf21}) and (\ref{mf27}) for the symmetrical body\footnote{The inertia tensor in this section is denoted by ${\mathbb I}$.}:  
${\mathbb I}=diagonal (I_2, I_2, I_3)$. This implies the following mass matrix: $g=\frac12 diagonal (I_3, I_3, 2I_2-I_3)$. We consider the positively charged body, then the charge-mass ratio is a positive number, $\mu>0$. 
We assume that at $t=0$ the third inertia axis of the body is vertical. Then, without spoiling the diagonal form of inertia tensor in our equations, the Laboratory system can be chosen as shown on 
Figure \ref{Lssc}.
The basis vector ${\bf e}_3$ is directed along the third inertia axis ${\bf R}_3(0)$, the vectors ${\bf e}_2$ and ${\bf e}_3$ lie on the plane of paper sheet together with the vector of constant magnetic field ${\bf B}=(0, B_2>0, B_3>0)^T$. The initial instantaneous angular velocity of the body is 
\begin{eqnarray}\label{mf41.0}
{\boldsymbol\Omega}(0)={\boldsymbol\omega}(0)=(\omega_1, \omega_2, \omega_3)^T. 
\end{eqnarray}
It is convenient to introduce the following notation:  
\begin{eqnarray}\label{mf41}
{\bf B}'\equiv\frac{\mu}{2c}{\bf B}=
\left(
\begin{array}{c}
0\\  B'_2>0 \\ B'_3>0 
\end{array}
\right), \qquad 
{\bf K}(t)\equiv\frac{\mu}{2c}{\bf B}R(t)=\left(
\begin{array}{c}
({\bf B}'R)_1 \\ ({\bf B}'R)_2 \\ ({\bf B}'R)_3 
\end{array}
\right), \qquad \mbox{then} \quad {\bf K}(0)={\bf B}';  
\end{eqnarray}
\begin{eqnarray}\label{mf41.1}
I\equiv\frac{I_3}{I_2}, \qquad E'\equiv\frac{2E}{I_2}.  
\end{eqnarray}
Contracting the Poisson equations (\ref{mf21}) with $B'_i$, we get $\dot{\bf K}=-[{\boldsymbol\Omega}, {\bf K}]$. This equation together with (\ref{mf27}) give us the auxiliary system of $3+3$ closed equations for determining the variables ${\boldsymbol\Omega}(t)$ and ${\bf K}(t)$  
\begin{eqnarray}
{\mathbb I}\dot{\boldsymbol\Omega}=[{\mathbb I}{\boldsymbol\Omega}, {\boldsymbol\Omega}]-2[{\bf K}g, {\boldsymbol\Omega}], \label{mf42} \\
\dot{\bf K}=-[{\boldsymbol\Omega}, {\bf K}]. \qquad \qquad   \label{mf43}
\end{eqnarray}
By construction, the initial conditions for ${\bf K}(t)$ are ${\bf K}(0)={\bf B}'$. 
Any solution $R_{ij}(t)$, $\Omega_i(t)$ to Euler-Poisson equations obeys to this system. So we can use the latter to look for the angular velocity $\Omega_i(t)$. 

This system admits four integrals of motion. Two of them are 
\begin{eqnarray}
\Omega_1^2+\Omega_2^2+I\Omega_3^2=E'=\omega_1^2+\omega_2^2+I\omega_3^2, \label{mf44} \\
{\bf K}^2=B'^2_2+B'^2_3. \qquad \qquad   \label{mf45}
\end{eqnarray}
To obtain two more integrals, we write our system in components 
\begin{eqnarray}\label{mf46}
\dot\Omega_1=(1-I)\Omega_3\Omega_2-I\Omega_3 K_2+(2-I)\Omega_2 K_3, \quad \cr 
\dot\Omega_2=-(1-I)\Omega_3\Omega_1+I\Omega_3 K_1-(2-I)\Omega_1 K_3, \cr 
\dot\Omega_3=\Omega_1 K_2-\Omega_2 K_1; \qquad \qquad \qquad \qquad \qquad \quad 
\end{eqnarray}
\begin{eqnarray}\label{mf47}
\dot K_1=\Omega_3 K_2-\Omega_2 K_3, \quad  \cr 
\dot K_2=-\Omega_3 K_1+\Omega_1 K_3, ~  \cr 
\dot K_3=-(\Omega_1 K_2-\Omega_2 K_1).  \label{mf47}
\end{eqnarray}
The equations with $\dot\Omega_3$ and $\dot K_3$ imply the third integral
\begin{eqnarray}\label{mf48}
K_3+\Omega_3=c_3=\omega_3+B'_3.
\end{eqnarray}
Combining the equations with $\dot\Omega_1$, $\dot\Omega_2$, $\dot K_1$ and $\dot K_2$ we get one more integral of motion
\begin{eqnarray}\label{mf49}
\Omega_1 K_1+\Omega_2 K_2-\Omega_3^2+(2-I)c_3\Omega_3=c_5=\omega_2 B'_2-\omega_3^2+(2-I)(\omega_3+B'_3)\omega_3. 
\end{eqnarray}
We written them through the integrations constants $c_3$ and $c_5$, as well as through the initial data $\omega_i$ and $B'_i$ of the problem. 

Using (\ref{mf48}), (\ref{mf49}) and the equation with $\dot\Omega_3$ of the system (\ref{mf46}), we represent the variables $K_i$ through $\Omega_i$ as follows
\begin{eqnarray}\label{mf50}
K_1=\frac{\Omega_1(\Omega_3^2-(2-I)c_3\Omega_3+c_5)-\Omega_2\dot\Omega_3}{E'-I\Omega_3^2}, \quad
K_2=\frac{\Omega_2(\Omega_3^2-(2-I)c_3\Omega_3+c_5)+\Omega_1\dot\Omega_3}{E'-I\Omega_3^2}, \quad 
K_3=-\Omega_3+c_3.
\end{eqnarray}
Substituting them into equations for $\dot\Omega_1$ and $\dot\Omega_2$ from (\ref{mf46}), we get 
\begin{eqnarray}\label{mf51}
\dot\Omega_1=\phi(t)\Omega_2+\dot f(t)\Omega_1,  \qquad \dot\Omega_2=-\phi(t)\Omega_1+\dot f(t)\Omega_2,
\end{eqnarray}
where $\phi(t)$ and $f(t)$ turn out to be the following functions of $\Omega_3(t)$:
\begin{eqnarray}\label{mf52}
\phi(t)\equiv\frac{(2-I)c_3E'-(E'+Ic_5)\Omega_3(t)}{E'-I\Omega_3^2(t)}, \qquad 
f(t)=\ln\sqrt{E'-I\Omega_3^2(t)}.
\end{eqnarray}
If $\Omega_3(t)$ is known, the equations (\ref{mf51}) can be immediately integrated as follows
\begin{eqnarray}\label{mf53}
\Omega_1(t)=f_0\sqrt{E'-I\Omega_3^2(t)}\sin(\Phi(t)+\phi_0), \qquad  \Omega_2(t)=f_0\sqrt{E'-I\Omega_3^2(t)}\cos(\Phi(t)+\phi_0),
\end{eqnarray}
where $\Phi(t)$ is indefinite integral of $\phi(t)$, while $f_0$ and $\phi_0$ are the integration constants. 

So it remains to find the third component $\Omega_3(t)$. To this aim we compute the time derivative of the last equation from (\ref{mf46}), and use other equations of the system (\ref{mf46}), (\ref{mf47}) in the resulting expression, presenting it as follows 
\begin{eqnarray}\label{mf54}
\ddot\Omega_3=[-2\Omega_3+(2-I)c_3](\Omega_1 K_1+\Omega_2 k_2)-I\Omega_3(K_1^2+K_2^2)+[c_3-\Omega_3](\Omega_1^2+\Omega_2^2). 
\end{eqnarray}
Using the integrals of motion (\ref{mf44}), (\ref{mf45}) and (\ref{mf49}), we obtain closed equation for determining $\Omega_3(t)$, that can be called the equation of cubic pseudo-oscillator
\begin{eqnarray}\label{mf55}
\ddot\Omega_3=a_0+a_1\Omega_3+a_2\Omega_3^2+a_3\Omega_3^3, 
\end{eqnarray}
where the numeric coefficients $a_i$ are functions of initial data of original problem
\begin{eqnarray}\label{mf56}
a_0({\boldsymbol\omega}, {\bf B}')=[(2-I)c_5+ E']c_3, \qquad 
a_1({\boldsymbol\omega}, {\bf B}')=-2c_5+[I-(2-1)^2]c_3^2-E'-I{\bf B}'^2, \cr 
a_2({\boldsymbol\omega}, {\bf B}')=6(1-I)c_3, \qquad a_3=-2(1-I). \qquad \qquad \qquad \qquad \qquad 
\end{eqnarray}
It is not difficult to obtain a two-parametric family of simple solutions to the equation (\ref{mf55}).
Note that $\Omega_3(t)=\omega_3$ will be (constant) solution to (\ref{mf55}) if the third component $\omega_3$ if initial angular velocity  is a root of the third degree polynomial on the right side of (\ref{mf55}).  Substituting $\Omega_3(t)=\omega_3$ into Eq. (\ref{mf55}), we obtain the condition on initial 
data $\omega_1, \omega_2, \omega_3$ under which $\omega_3$ satisfies this equation. To obtain this condition, it is convenient to represent Eq. (\ref{mf55}) in terms of the initial data, keeping the 
combinations like $\omega_3-\Omega_3(t)$ as follows:
\begin{eqnarray}\label{mf57}
\ddot\Omega_3=[(2-I)B'_3-I\omega_3+2(\omega_3-\Omega_3)][\omega_2B'_2-(\omega_3^2-\Omega_3^2)+(2-I)(B'_3+\omega_3)(\omega_3-\Omega_3)]- \cr
I\Omega_3[{\bf B}'^2-(B'_3+(\omega_3-\Omega_3))^2]+[B'_3+(\omega_3-\Omega_3)][\omega_1^2+\omega_2^2+I(\omega_3^2-\Omega_3^2)]. \qquad \qquad 
\end{eqnarray}
Substituting $\Omega_3(t)=\omega_3$, we get that (\ref{mf55}) will be satisfied only for the initial data $\omega_i$ obeying the following equation: 
\begin{eqnarray}\label{mf58.1}
B'_3(\omega_1^2+\omega_2^2)-IB'_2\omega_2\omega_3+(2-I)B'_2B'_3\omega_2-IB'^2_2\omega_3=0,
\end{eqnarray}
or, equivalently
\begin{eqnarray}\label{mf58}
\omega_1^2+(\omega_2+B'_2)^2-\frac{IB'_2}{B'_3}(\omega_2+B'_2)(\omega_3+B'_3)=(1-I)B'^2_2. 
\end{eqnarray}
This is a surface of second order. 
Since the point with $\omega_i=0$ obeys this equation, the surface always pass through the origin of coordinate system. Resolving (\ref{mf58.1}) with respect to $\omega_3$, we get the following  two-parametric family of constant solutions to the equation (\ref{mf55}) of cubic pseudo-oscillator: 
\begin{eqnarray}\label{mf59}
\Omega_3(t)=\omega_3(\omega_1, \omega_2)=\frac{B'_3[\omega_1^2+\omega_2^2+(2-I)B' _2\omega_2]}{IB'_2(\omega_2+B'_2)}, \qquad 
\omega_1\in{\mathbb R}, \quad 
\omega_2\in{\mathbb R}\setminus \{-B'_2\}. 
\end{eqnarray}

\begin{figure}[t] \centering
\includegraphics[width=12cm]{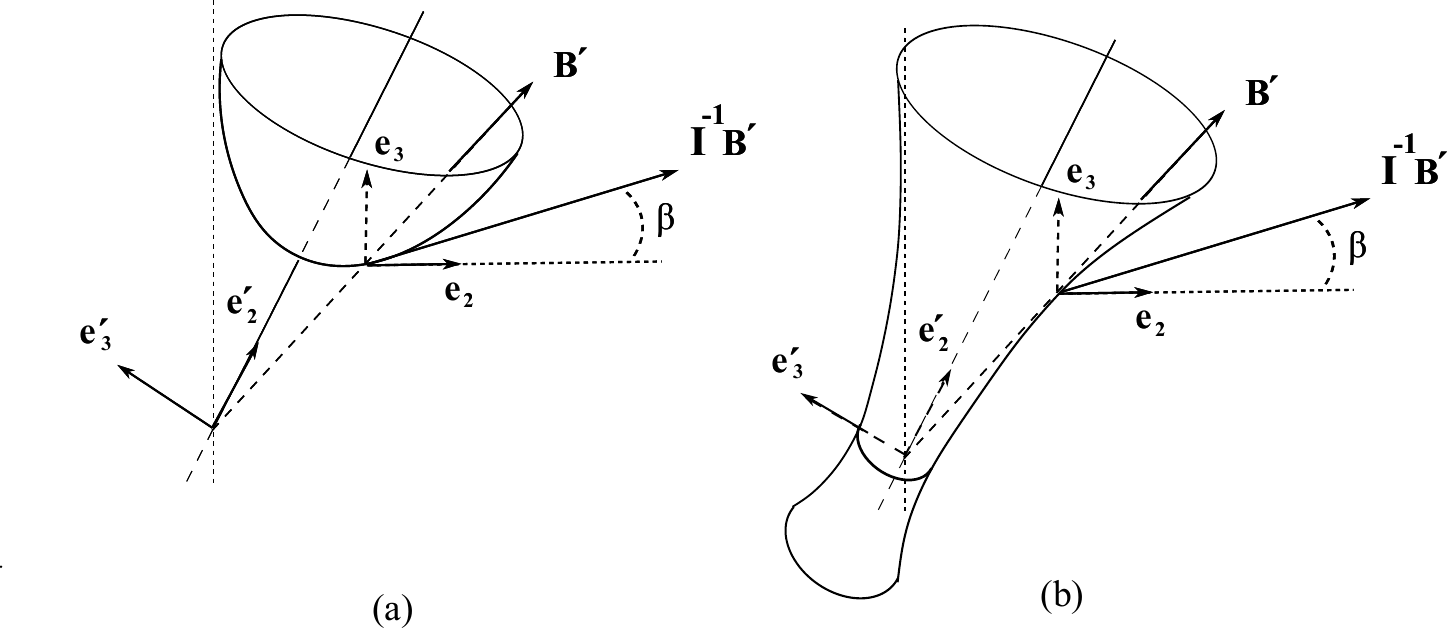}
\caption{The third components of initial data vectors ${\boldsymbol\omega}$, lying on drawn hyperboloids, turn out to be solutions to the equation of cubic pseudo-oscillator (\ref{mf55}). Figure (a) shows the upper sheet of the hyperboloid of a short solenoid. Figure (b) shows the hyperboloid of a long solenoid. All drawn vectors and line segments lie in the plane of paper sheet. The basis vector ${\bf e}_1$ is orthogonal to the plane of paper sheet and is not shown in the figure.}\label{Hyp}
\end{figure}

Let us find out which quadric is defined by the equation (\ref{mf58}), by writing it in the canonical form. Following the standard 
procedure \cite{Shi_1977}, we arrive at the new coordinates $\omega'_i$:
\begin{eqnarray}\label{mf61}
\omega'_1=\omega_1, \qquad \qquad \qquad \qquad \qquad \qquad \qquad \qquad \qquad \quad \qquad \cr
\omega'_2=(\omega_2+B'_2)\cos\left(\frac{\beta}{2}+\frac{\pi}{4}\right)+(\omega_3+B'_3)\sin\left(\frac{\beta}{2}+\frac{\pi}{4}\right), \cr
\omega'_3=-(\omega_2+B'_2)\sin\left(\frac{\beta}{2}+\frac{\pi}{4}\right)+(\omega_3+B'_3)\cos\left(\frac{\beta}{2}+\frac{\pi}{4}\right), 
\end{eqnarray}
where $\beta$ is the angle between the vectors ${\bf e}_2$ and ${\mathbb I}^{-1}{\bf B}'$, see Figure \ref{Hyp}.
The new coordinates are obtained from $\omega_i$ by shifting the origin of coordinate system to the point $-{\bf B}'$, and subsequent rotation counter-clockwise by the angle $\frac{\beta}{2}+\frac{\pi}{4}$ in the plane ${\bf e}_2, {\bf e}_3$. 
Note that $\frac{\pi}{4}<\frac{\beta}{2}+\frac{\pi}{4}<\frac{\pi}{2}$. In these coordinates  Eq. (\ref{mf58}) acquires the form 
\begin{eqnarray}\label{mf60}
\omega'^2_1-\frac{\omega'^2_2}{a_2}+\frac{\omega'^2_3}{a_3}=(1-I)B'^2_2, \qquad \mbox{where} \quad a_2\equiv\frac{2\sin\beta}{1-\sin\beta}, \quad a_3\equiv\frac{2\sin\beta}{1+\sin\beta}.
\end{eqnarray}
Depending on the relationship between the inertia moments $I_2$ and $I_3$, it describes different surfaces. \par

\noindent {\bf 1.} Let $1<I\equiv\frac{I_3}{I_2}<2$. This body could be a charged sufficiently short cylindrical surface. If it rotates around its coaxial axis, it will produce a magnetic field corresponding to a short solenoid. The equation (\ref{mf60}) turn into 
\begin{eqnarray}\label{mf61}
\frac{\omega'^2_1}{C_1^2}-\frac{\omega'^2_2}{C_2^2}+\frac{\omega'^2_3}{C_3^2}=-1, \qquad \mbox{where} \quad 
C_1^2\equiv(I-1)B'^2_2, \quad 
C_2^2\equiv\frac{2(I-1)B'^2_2\sin\beta}{1-\sin\beta}, \quad C_3^2\equiv\frac{2(I-1)B'^2_2\sin\beta}{1+\sin\beta}.
\end{eqnarray}
Hence the surface of initial data is a hyperboloid of two sheets. Its upper sheet is shown in Figure \ref{Hyp} (a). 

In the limiting case $I\equiv\frac{I_3}{I_2}=2$ we have a plane body, that could be charged circular loop.  In this case the sheets of the hyperboloid are tangent to the horizontal planes $\omega_3=0$ and $\omega_3=-2B'_3$. \par 

\noindent {\bf 2.} For the totally symmetric body $I\equiv\frac{I_3}{I_2}=1$, the equation (\ref{mf60}) turn into the cone
\begin{eqnarray}\label{mf62}
\omega'^2_1+\frac{\omega'^2_3}{a_3}=\frac{\omega'^2_2}{a_2},  
\end{eqnarray}
with semi axes $a_2$ and $a_3$ written in equation (\ref{mf60}). 
\par

\noindent {\bf 3.} Let $0<I\equiv\frac{I_3}{I_2}<1$. This body could be a charged long cylindrical surface. If it rotates around its coaxial axis, it will produce a magnetic field corresponding to a long solenoid. The equation (\ref{mf60}) turn into 
\begin{eqnarray}\label{mf63}
\frac{\omega'^2_1}{C_1^2}-\frac{\omega'^2_2}{C_2^2}+\frac{\omega'^2_3}{C_3^2}=1, \qquad \mbox{where} \quad 
C_1^2\equiv(1-I)B'^2_2, \quad 
C_2^2\equiv\frac{2(1-I)B'^2_2\sin\beta}{1-\sin\beta}, \quad C_3^2\equiv\frac{2(1-I)B'^2_2\sin\beta}{1+\sin\beta}.
\end{eqnarray}
Hence the surface of initial data is a hyperboloid of one sheet shown in Figure \ref{Hyp} (b). 

In resume, we have shown that for any solution to the Euler-Poisson equations (\ref{mf21}) and (\ref{mf27}) of a symmetrical charged body, the 
function $\Omega_3(t)$ obeys the equation of cubic pseudo-oscillator (\ref{mf55}).  We obtained a two-parameter family of constant solutions (\ref{mf59}) to this equation. Not all of them generate solutions to the original problem. In the next section they will help us to obtain  a one-parameter family of solutions to the original Euler-Poisson equations in elementary functions.

\section{Rotation matrix: one-parametric family of solutions in elementary functions.}\label{CH6.6}  

As we saw in previous section, our problem (\ref{mf21}), (\ref{mf27}) probably admits  solutions with constant $\Omega_3(t)$. So let us search for solutions of the auxiliary task (\ref{mf42}), (\ref{mf43}) of the form 
\begin{eqnarray}\label{mf64}
{\boldsymbol\Omega}(t)=(\Omega_1(t), ~ \Omega_2(t), ~ \omega_3=\mbox{const})^T, \qquad 
{\boldsymbol\Omega}(0)=(\omega_1, \omega_2, \omega_3)^T. 
\end{eqnarray}
Substituting this ansatz into the equations (\ref{mf42}) and (\ref{mf43}), they turn into 
\begin{eqnarray}\label{mf65}
\dot\Omega_1=[(1-I)\omega_3+(2-I)B'_3]\Omega_2-I\omega_3K_2, \quad 
\dot\Omega_2=-[(1-I)\omega_3+(2-I)B'_3]\Omega_1+I\omega_3K_1, \quad 
\Omega_1K_2-\Omega_2K_1=0; 
\end{eqnarray}
\begin{eqnarray}\label{mf66}
\dot K_1=\omega_3K_2-B'_3\Omega_2, \qquad \dot K_2=-\omega_3K_1+B'_3\Omega_1, \qquad K_3=B'_3.
\end{eqnarray}
These equations admit three integrals of motion:  $\frac{d}{dt}(\Omega_1^2+\Omega_2^2)=0$, $\frac{d}{dt}(K_1^2+K_2^2)=0$ 
and $\frac{d}{dt}(\Omega_1K_1+\Omega_2K_2)=0$. This implies the equalities
\begin{eqnarray}\label{mf67}
\Omega_1^2(t)+\Omega_2^2(t)=\omega_1^2+\omega_2^2, \qquad 
K_1^2(t)+K_2^2(t)=B'^2_2, \qquad \Omega_1(t)K_1(t)+\Omega_2(t)K_2(t)=\omega_2B'_2. 
\end{eqnarray}
Using the equations $\Omega_1K_1+\Omega_2K_2=\omega_2B'_2$ and $-\Omega_2K_1+\Omega_1K_2=0$ we get that $K_1$ and $K_2$ are just proportional to $\Omega_1$ and $\Omega_2$
\begin{eqnarray}\label{mf68}
K_1=\frac{\omega_2B'_2}{\omega_1^2+\omega_2^2}\Omega_1, \qquad K_2=\frac{\omega_2B'_2}{\omega_1^2+\omega_2^2}\Omega_2. 
\end{eqnarray}
Substituting these expressions into (\ref{mf65}) and (\ref{mf66}), we get the equations of precession 
\begin{eqnarray}\label{mf69}
\dot\Omega_1=\phi'\Omega_2, \qquad  \dot\Omega_2=-\phi'\Omega_1, \qquad \mbox{where} \quad \phi'\equiv(1-I)\omega_3+(2-I)B'_3-\frac{IB'_2\omega_2\omega_3}{\omega_1^2+\omega_2^2}; 
\end{eqnarray}
and 
\begin{eqnarray}\label{mf70}
\dot\Omega_1=\phi\Omega_2, \qquad  \dot\Omega_2=-\phi\Omega_1, \qquad \mbox{where} \quad 
\phi\equiv \omega_3-\frac{B'_3(\omega_1^2+\omega_2^2)}{B'_2\omega_2}. \qquad \qquad \qquad \quad 
\end{eqnarray}
They will be consistent only if $\phi'=\phi$, that is the initial data should lie on the surface
\begin{eqnarray}\label{mf71}
B'_3(\omega_1^2+\omega_2^2)-IB'_2\omega_2\omega_3+(2-I)B'_2B'_3\omega_2-\frac{IB'^2_2\omega^2_2\omega_3}{\omega_1^2+\omega_2^2}=0.
\end{eqnarray}
Combining this with the necessary condition (\ref{mf58.1})
\begin{eqnarray}\label{mf71.1}
B'_3(\omega_1^2+\omega_2^2)-IB'_2\omega_2\omega_3+(2-I)B'_2B'_3\omega_2-IB'^2_2\omega_3=0,
\end{eqnarray}
we conclude that the initial data should be taken on the curve of second-order 
\begin{eqnarray}\label{mf72}
B'_3\omega_2^2-IB'_2\omega_2\omega_3+(2-I)B'_2B'_3\omega_2-IB'^2_2\omega_3=0,
\end{eqnarray}
that lie on the plane $\omega_1=0$. Geometrically, these are hyperbolas that appear as a result of the intersection of the hyperboloids in 
Figure \ref{Hyp} with this plane. 

For the circular loop or short solenoid they are 
\begin{eqnarray}\label{mf73}
\frac{\omega'^2_2}{C_2^2}-\frac{\omega'^2_3}{C_3^2}=1, \qquad \mbox{where} \quad  
C_2^2\equiv\frac{2(I-1)B'^2_2\sin\beta}{1-\sin\beta}, \quad C_3^2\equiv\frac{2(I-1)B'^2_2\sin\beta}{1+\sin\beta}.
\end{eqnarray}
For the long solenoid they are
\begin{eqnarray}\label{mf74}
-\frac{\omega'^2_2}{C_2^2}+\frac{\omega'^2_3}{C_3^2}=1, \qquad \mbox{where} \quad  
C_2^2\equiv\frac{2(1-I)B'^2_2\sin\beta}{1-\sin\beta}, \quad C_3^2\equiv\frac{2(1-I)B'^2_2\sin\beta}{1+\sin\beta}.
\end{eqnarray}
At last, for a totally symmetric body they degenerate into the straight lines
\begin{eqnarray}\label{mf75}
\frac{\omega'_2}{a_2}=\pm\frac{\omega'_3}{a_3}, \qquad \mbox{where} \quad  
a_2\equiv\sqrt{\frac{2\sin\beta}{1-\sin\beta}}, \quad a_3\equiv\sqrt{\frac{2\sin\beta}{1+\sin\beta}}.
\end{eqnarray}

Resolving Eq. (\ref{mf72}) with respect to $\omega_3$ we get
\begin{eqnarray}\label{mf76}
\omega_3(\omega_2)=\frac{B'_3\omega_2[\omega_2+(2-I)B'_2]}{IB'_2(\omega_2+B'_2)}, \qquad 
\omega_2\in{\mathbb R}\setminus\{-B'_2\}.
\end{eqnarray}
With this $\omega_3$, the two systems (\ref{mf69}) and (\ref{mf70}) depend on the same frequency
\begin{eqnarray}\label{mf77}
\phi=\omega_3(\omega_2)-\frac{B'_3}{B'_2}\omega_2=\frac{B'_3\omega_2(1-I)(\omega_2+2B'_2)}{IB'_2(\omega_2+B'_2)},  
\end{eqnarray}
and imply the following solution 
\begin{eqnarray}\label{mf78}
{\boldsymbol\Omega}(t)=\left(
\begin{array}{c}
\omega_2\sin\phi t \\ \omega_2\cos\phi t \\ \omega_3(\omega_2)
\end{array}
\right), \qquad 
\mbox{then} \quad {\boldsymbol\Omega}(0)=\left( \begin{array}{c} 0 \\ \omega_2 \\ \omega_3(\omega_2) \end{array}\right). 
\end{eqnarray}
This means that vector of angular velocity in the body-fixed frame ${\bf R}_i$ precesses around the third axis ${\bf R}_3$ with the 
frequency $\phi$. 

The next step is to solve the Poisson equations (\ref{mf21}). We consider them in the form  
\begin{eqnarray}\label{mf79}
\dot{\boldsymbol\gamma}=[{\boldsymbol\gamma}, {\boldsymbol\Omega}(t)],
\end{eqnarray}
with ${\boldsymbol\Omega}(t)$ specified by  Eqs. (\ref{mf76})-(\ref{mf78}). Here ${\boldsymbol\gamma}$ is any one of rows of the rotation matrix.

In components this reads
\begin{eqnarray}\label{mf80}
\dot\gamma_1=\gamma_2\omega_3-\gamma_3\Omega_2(t), \qquad 
\dot\gamma_2=\gamma_3\Omega_1(t)-\gamma_1\omega_3, \qquad 
\dot\gamma_3=\gamma_1\Omega_2(t)-\gamma_2\Omega_1(t). \qquad 
\end{eqnarray}
This system admits the integral of motion 
\begin{eqnarray}\label{mf81}
\Omega_1\gamma_1+\Omega_2\gamma_2+\frac{\hat B_3\omega_2}{\hat B_2}\gamma_3=c,
\end{eqnarray}
where $\hat B_2\equiv B_2/|{\bf B}|$ and $\hat B_3\equiv B_3/|{\bf B}|$ are components of unit vector in the direction of magnetic vector ${\bf B}$.
Preservation in time of the quantity (\ref{mf81}) can  be verified by direct computation of its time-derivative, with use the 
identities  
\begin{eqnarray}\label{mf82}
\frac{B'_3}{B'_2}=\frac{B_3}{B_2}=\frac{\hat B_3}{\hat B_2}. 
\end{eqnarray}

We need to find the general solution to the equations (\ref{mf80}). Then, 
according to \cite{AAD23},  the rows of the rotation matrix $R_{ij}$ can be obtained taking the following three particular solutions.  The row $R_{11}, R_{12}, R_{13}$ is ${\boldsymbol\gamma}(t)$ with the initial data ${\boldsymbol\gamma}(0)=(1, 0, 0)$ and with $c=\Omega_1(0)=0$. The row $R_{21}, R_{22}, R_{23}$ is ${\boldsymbol\gamma}(t)$ with the initial data ${\boldsymbol\gamma}(0)=(0, 1, 0)$ and with $c=\Omega_2(0)=\omega_2$. At last, the row $R_{31}, R_{32}, R_{33}$ is ${\boldsymbol\gamma}(t)$ with the initial data ${\boldsymbol\gamma}(0)=(0, 0, 1)$ and 
with $c=\hat B_3\omega_2/\hat B_2$.

First we solve algebraically the equations $\Omega_1\gamma_1+\Omega_2\gamma_2=c-\frac{\hat B_3\omega_2}{\hat B_2}\gamma_3$ 
and $\Omega_2\gamma_1-\Omega_1\gamma_2=\dot\gamma_3$, representing $\gamma_1$ and $\gamma_2$ as follows:
\begin{eqnarray}\label{mf83}
\gamma_1=\frac{1}{\omega_2^2}[\Omega_1(c-\frac{\hat B_3\omega_2}{\hat B_2}\gamma_3)+\Omega_2\dot\gamma_3], \qquad 
\gamma_2=\frac{1}{\omega_2^2}[\Omega_2(c-\frac{\hat B_3\omega_2}{\hat B_2}\gamma_3)-\Omega_1\dot\gamma_3],
\end{eqnarray}
Substituting them into the equation for $\dot\gamma_1$, we get closed equation of second order for $\gamma_3(t)$ 
\begin{eqnarray}\label{mf84}
\ddot\gamma_3+k^2\gamma_3=c\frac{\hat B_3\omega_2}{\hat B_2}, \qquad \mbox{where} \quad k=\frac{\omega_2}{\hat B_2}. 
\end{eqnarray}
This is equation of harmonic oscillator with constant frequency $k$, under the action of an external constant force. Its general solution with the integration constants $b$ and $k_0$ is 
\begin{eqnarray}\label{mf85}
\gamma_3(t)=b\cos(kt+k_0)+c\frac{\hat B_3\omega_2}{\hat B_2 k^2}. 
\end{eqnarray}
Substituting this result into the expressions (\ref{mf83}), we obtain the remaining variables
\begin{eqnarray}\label{mf86}
\gamma_1(t)=\frac{1}{\omega_2}[c-b\frac{\hat B_3\omega_2}{\hat B_2}\cos(kt+k_0)-
c\hat B_3^2]\sin\phi t-\frac{b}{\hat B_2}\sin(kt+k_0)\cos\phi t, \cr
\gamma_2(t)=\frac{1}{\omega_2}[c-b\frac{\hat B_3\omega_2}{\hat B_2}\cos(kt+k_0)-
c\hat B_3^2]\cos\phi t+\frac{b}{\hat B_2}\sin(kt+k_0)\sin\phi t.
\end{eqnarray}
At $t=0$ we get
\begin{eqnarray}\label{mf87}
\gamma_1(0)=-\frac{b}{\hat B_2}\sin k_0, \qquad \gamma_2(0)=\frac{1}{\omega_2}[c-b\frac{\hat B_3\omega_2}{\hat B_2}\cos k_0-
c\hat B_3^2], \qquad \gamma_3(0)=b\cos k_0+c\frac{\hat B_2\hat B_3}{\omega_2}. 
\end{eqnarray}
Solving the equations (\ref{mf87}) with the data described below Eq. (\ref{mf82}), we get, in each case
\begin{eqnarray}\label{mf88}
c=0, \quad b=-\hat B_2, \quad k_0=\frac{\pi}{2}; \qquad  c=\omega_2, \quad b=-\hat B_2\hat B_3, \quad k_0=0; \qquad 
c=\frac{\hat B_3\omega_2}{\hat B_2}, \quad b=1-\hat B_3^2=\hat B_2^2, \quad k_0=0.
\end{eqnarray}
Substituting these values into Eqs. (\ref{mf85}) and (\ref{mf86}) we get the rotation matrix of a symmetrical charged body, immersed into the magnetic field ${\bf B}=(0, B_2, B_3)^T$, and launched with initial angular velocity (\ref{mf78}) 
\begin{eqnarray}\label{mf89}
\left(
\begin{array}{ccc}
\cos kt\cos\phi t-\hat B_3\sin kt\sin\phi t & -\cos kt\sin\phi t-\hat B_3\sin kt\cos\phi t  &  \hat B_2\sin kt  \\
{} & {} & {} \\
\hat B_3 \sin kt\cos\phi t+(\hat B_2^2 +\hat B_3^2\cos kt)\sin\phi t & 
-\hat B_3 \sin kt\sin\phi t+(\hat B_2^2 +\hat B_3^2\cos kt)\cos\phi t & \hat B_2\hat B_3(1-\cos kt) \\
{} & {} & {} \\
-\hat B_2 \sin kt\cos\phi t+\hat B_2 \hat B_3(1-\cos kt)\sin\phi t &
\hat B_2 \sin kt\sin\phi t+\hat B_2 \hat B_3(1-\cos kt)\cos\phi t & \hat B_3^2 +\hat B_2^2\cos kt
\end{array}\right) 
\end{eqnarray}
Two frequencies in the problem are: $\phi$ written in Eq. (\ref{mf77}), and $k=\omega_2/\hat B_2$. The dependence of the rotation matrix on the inertia moments $I_2$, $I_3$ as well as on the charge-mass ratio $\mu$ is hidden in the frequency $\phi$. 

By direct substitution of obtained functions (\ref{mf78}) and (\ref{mf89})  into the equations (\ref{mf21}) and (\ref{mf27}), I verified that they are satisfied.

The rotation matrix  can be decomposed as follows: 
\begin{eqnarray}\label{mf90}
R(t)=R_{\hat {\bf B}}(t, k)\times R_{OZ}(t, \phi)=
\left(
\begin{array}{ccc}
\cos kt & -\hat B_3\sin kt  &  \hat B_2\sin kt  \\
{} & {} & {} \\
\hat B_3 \sin kt & 
\hat B_2^2 +\hat B_3^2\cos kt & \hat B_2\hat B_3(1-\cos kt) \\
{} & {} & {} \\
-\hat B_2 \sin kt & \hat B_2 \hat B_3(1-\cos kt) & \hat B_3^2 +\hat B_2^2\cos kt
\end{array}\right)\times
\left(
\begin{array}{ccc}
\cos \phi t  & -\sin \phi t & 0  \\ 
\sin \phi t &  \cos \phi t & 0 \\
0 & 0 & 1 \\
\end{array}\right) 
\end{eqnarray}
Then the position ${\bf x}(t)$ of any point of the body at the 
instant $t$ is: ${\bf x}(t)=R_{\hat{\bf B}}(t, k)\times R_{OZ}(t, \phi){\bf x}(0)$. It is obtained by rotating the initial position vector ${\bf x}(0)$ first around the laboratory axis $OZ$ by the angle $\phi t$ and then around the ${\bf B}$\,-axis by the angle $kt$. 

It can be said that the motion is the composition of a proper rotation around third inertia axis with precession of this axis around the vector of magnetic field ${\bf B}$. The final answer (\ref{mf89}) admits the limit of totally symmetric body $I_2=I_3$, this implies $\phi=0$. The resulting motion is the precession around the magnetic vector ${\bf B}$ without a proper rotation. 

Combining Eqs. (\ref{mf76}), (\ref{mf77}) and (\ref{mf84}) we get the relation between two frequencies of the motion (\ref{mf89})
\begin{eqnarray}\label{mf91}
\phi_B=\frac{(I_2-I_3)\hat B_3(2|{\bf B}'|+k_B)}{I_3(|{\bf B}'|+k_B)}k_B. 
\end{eqnarray}

We recall that the most general motion of a free symmetrical body is the precession without nutation \cite{AAD23}. Observe that the rotation matrix (\ref{mf89}) coincides with Eq. (132) of this work if we replace $\hat {\bf B}$, $k_B$ and $\phi_B$ on $\hat {\bf m}$, $k$ and $\phi$. The physical meaning of this coincidence can be formulated as follows. \par

\noindent
{\bf Affirmation.} If a symmetrical charged body in the magnetic field ${\bf B}=(0, B_2, B_3)$ moves according (\ref{mf89}) with the precession frequency $k_B$ around ${\bf B}$ and the proper rotation frequency $\phi_B$, then in the absence of a magnetic field its precession with the same frequency $k_B$ around the unit vector $\hat {\bf B}$  will happen with the proper rotation frequency 
\begin{eqnarray}\label{mf92}
\phi=\frac{I_2-I_3}{I_3}\hat B_3k_B. 
\end{eqnarray}
Indeed, consider the motion (\ref{mf89}) with initial angular velocity ${\boldsymbol\omega}(0)=(0, \omega_2, \omega_3(\omega_2))^T$. Let it then was launched in the absence of a magnetic field with initial angular 
velocity ${\boldsymbol\Omega}(0)=(0, \omega_2, \frac{I_2\omega_2B_3}{I_3B_2})^T$. 
According to \cite{AAD23},  it will precess around the vector of conserved angular 
momentum ${\bf m}=I{\boldsymbol\Omega}(0)=(0, I_2\omega_2, \frac{I_2\omega_2}{B_2}B_3)=\frac{I_2\omega_2}{B_2}{\bf B}\sim {\bf B}$ with the frequency $k=\frac{|{\bf m}|}{I_2}=\frac{\omega_2}{\hat B_2}=k_B$ and with the proper rotation 
frequency $\phi=\frac{I_2-I_3}{I_2I_3}m_3=\frac{I_2-I_3}{I_3}\hat B_3\frac{\omega_2}{\hat B_2}=\frac{I_2-I_3}{I_3}\hat B_3k_B$.  

Components of angular momentum ${\bf m}(t)$ for our solutions in elementary functions are not  conserved quantities. But using the integrals of motion (\ref{mf44}), (\ref{mf48})  and (\ref{mf49}) with $\Omega_3(t)=\omega_2=\mbox{const}$, we get 
\begin{eqnarray}\label{mf93}
({\bf m}(t), {\bf B})=\mbox{const}. 
\end{eqnarray}
That is the angular momentum always lies in the plane orthogonal to the constant vector of magnetic field.

\section{Conclusion.}

In this work we deduced equations of motion of a charged symmetrical body in external constant and homogeneous electric and magnetic fields starting from the variational problem (\ref{1.6})  and (\ref{mf15}), where the body is considered as a system of charged point particles subject to holonomic constraints. The final equations are written in terms of center-of mass-coordinate, rotation matrix and angular velocity. They are (\ref{mf17}), (\ref{mf21}) and (\ref{mf27}). According to them, rotational motion of the body does not perturb its translational motion and vice-versa. In particular, the center of mass obeys to Eq. (\ref{mf17}) and behaves as a point charged particle in the electromagnetic field. Besides, the electric field does not affect the rotational motion of the body. 

For the case of a totally symmetrical body (charged ball) we found general solution (\ref{mf38}), (\ref{mf39}) to the equations of motion. The resulting motion can be thought as a superposition of two rotations: the first around unit vector $\hat{\bf k}$ written in Eq. (\ref{mf39}), with the frequency $\gamma$ determined by initial values of angular velocity and Larmor's frequency; and the second around the axis of magnetic 
field ${\bf B}=(0, 0, |{\bf B}|)^T$ with the frequency $\alpha=\frac{\mu|{\bf B}|}{2c}$. The angular momentum vector ${\bf m}(t)$ precesses around the vector ${\bf B}$ with the Larmor's frequency $\alpha=\frac{\mu|{\bf B}|}{2c}$. 

Analysing the equations (\ref{mf21}) and (\ref{mf27}) for the case of a symmetrical charged top, we demonstrated that the task to find the components of angular velocity $\Omega_i(t)$ can be reduced to solving the equation of a one-dimensional cubic 
pseudo-oscillator (\ref{mf55}). We found a two-parametric family  of solutions (\ref{mf59}) to this equation. This helped us later find a one-parametric family of solutions (\ref{mf89}) for the rotation matrix of a symmetrical charged body, immersed into the magnetic field ${\bf B}=(0, B_2, B_3)^T$, and launched with initial angular velocity (\ref{mf78}). The resulting motions turn out to be the composition of a proper rotation around third inertia axis with precession of this axis around the vector of magnetic field ${\bf B}$.

\begin{acknowledgments}
The work has been supported by the Brazilian foundation CNPq (Conselho Nacional de
Desenvolvimento Cient\'ifico e Tecnol\'ogico - Brasil). 
\end{acknowledgments}

\section{Appendix. Charged particle in constant and homogeneous electric and magnetic fields.}\label{CH6.2}
Consider a particle with mass $m$ and charge $e$ ($e<0$ for the electron), moving subject to constant and homogeneous electric ${\bf E}$ and magnetic ${\bf B}$ fields. Its dynamics is governed by the Lorentz-force equation
\begin{eqnarray}\label{mf1}
m\ddot{\bf y}=e{\bf E}+\frac{e}{c}[\dot{\bf y}, {\bf B}], \qquad \mbox{or} \quad \ddot{\bf y}=\mu{\bf E}+\frac{\mu}{c}[\dot{\bf y}, {\bf B}].
\end{eqnarray}
The universal constant $c$ is the speed of light in vacuum, and charge to mass ratio we denoted by $\mu=\frac{e}{m}$. This equation can be solved in elementary functions \cite{Landau_9}. One integration can be performed directly, giving a first-order equation with three integration 
constants being components of the vector ${\bf C}$
\begin{eqnarray}\label{mf2}
\dot{\bf y}=\mu{\bf E}t+\frac{\mu}{c}[{\bf y}, {\bf B}]+{\bf C}. 
\end{eqnarray}
When ${\bf B}=0$, particle moves with the acceleration directed along ${\bf E}$: ${\bf y}(t)={\bf y}_0+{\bf C}t+\frac12 \mu{\bf E}t^2$. When ${\bf E}=0$,  particle moves along a helical line located on a cylinder, the axis of which is directed along the vector ${\bf B}$. 

\begin{figure}[t] \centering
\includegraphics[width=08cm]{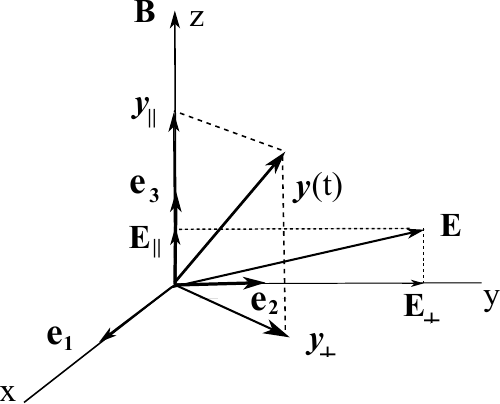}
\caption{Choice of Laboratory system for analysis of charged particle subject to electric and magnetic fields.}\label{Magn}
\end{figure}

To find an explicit form of the solution when both ${\bf E}$ and ${\bf B}$ are presented, we choose the Laboratory system with axis $z$ in the direction of ${\bf B}$, so that the vector ${\bf E}$ lies in $z, y$ plane, see Fig. \ref{Magn}. Consider the problem in the variables ${\bf y}(t)={\bf y}_{\Vert}(t)+{\bf y}_{\bot}(t)$, where ${\bf y}_{\Vert}(t)$ is the projection of our particle on axis ${\bf B}$ while ${\bf y}_{\bot}(t)$ is the projection on the plane orthogonal to ${\bf B}$. Similarly, we separate ${\bf C}={\bf C}_{\Vert}+{\bf C}_{\bot}$ and ${\bf E}={\bf E}_{\Vert}+{\bf E}_{\bot}$. Substituting  these decompositions into Eq. (\ref{mf2}), we get separate equations for two projections
\begin{eqnarray}
\dot{\bf y}_{\Vert}={\bf C}_{\Vert}+\mu{\bf E}_{\Vert}t, \qquad \qquad ~\quad \label{mf3} \\
\dot{\bf y}_{\bot}=\frac{\mu}{c}[{\bf y}_{\bot}, {\bf B}]+{\bf C}_{\bot}+\mu{\bf E}_{\bot}t.  \label{mf4} 
\end{eqnarray}
Integrating Eq. (\ref{mf3}) we get 
\begin{eqnarray}\label{mf5}
{\bf y}_{\Vert}(t)={\bf y}_{\Vert 0}+{\bf C}_{\Vert}t+\frac 12\mu{\bf E}_{\Vert}t^2. 
\end{eqnarray}
So the point ${\bf y}_{\Vert}(t)$ moves with the acceleration $|\mu{\bf E}_{\bot}|$ along  the axis ${\bf B}$. 

To solve Eq. (\ref{mf4}), we use the identity ${\bf E}_{\bot}=-[[{\bf E}_{\bot}, {\bf B}], {\bf B}]/{\bf B}^2$, which holds for any two orthogonal vectors. Then  Eq. (\ref{mf4}) reads as follows
\begin{eqnarray}\label{mf6}
\dot{\bf y}_{\bot}=\frac{\mu}{c}[{\bf y}_{\bot}-\frac{c t[{\bf E}_{\bot}, {\bf B}]}{|{\bf B}|^2}, {\bf B}]+{\bf C}_{\bot}.  
\end{eqnarray}
For the variable ${\bf y}_1\equiv {\bf y}_{\bot}-\frac{c t[{\bf E}_{\bot}, {\bf B}]}{{\bf B}^2}$ this equation implies
\begin{eqnarray}\label{mf7}
\dot{\bf y}_1=\frac{\mu}{c}[{\bf y}_1, {\bf B}]+{\bf C}_{\bot}-\frac{c [{\bf E}_{\bot}, {\bf B}]}{{\bf B}^2}, \qquad \mbox{or} \quad 
\dot{\bf y}_1=\frac{\mu}{c}[{\bf y}_1-\frac{c[{\bf C}_{\bot}, {\bf B}]+c^2{\bf E}_{\bot}}{\mu{\bf B}^2}, {\bf B}]. 
\end{eqnarray}
Then for the variable ${\bf y}_2\equiv {\bf y}_1-\frac{c[{\bf C}_{\bot}, {\bf B}]+c^2{\bf E}_{\bot}}{\mu{\bf B}^2}$ this equation implies 
\begin{eqnarray}\label{mf8}
\dot{\bf y}_2=-\frac{\mu}{c}[{\bf B}, {\bf y}_2],
\end{eqnarray}
that is ${\bf y}_2$ precesses on the plane $x, y$ around the vector ${\bf B}$. General solution to this equation is 
\begin{eqnarray}\label{mf9}
{\bf y}_2(t)=a\left\{{\bf e}_1\sin(\omega t+\omega_0)+{\bf e}_2\cos(\omega t+\omega_0)\right\}, \qquad \omega=\frac{|\mu{\bf B}|}{c}. 
\end{eqnarray}
where ${\bf e}_1$ and ${\bf e}_2$ are unit vectors in the direction of $x$ and $y$ axes. 
Returning back to the original variables, we get a general solution to Eq. (\ref{mf4}) with fourth integration constants ${\bf C}_{\bot}$, $a$ and $\omega_0$
\begin{eqnarray}\label{mf10}
{\bf y}_{\bot}(t)=\frac{[{\bf C}_{\bot}, {\bf B}]+c{\bf E}_{\bot}}{\omega|{\bf B}|}+\frac{\mu |{\bf E}_{\bot}|}{\omega}t{\bf e}_1+
a\left\{{\bf e}_1\sin(\omega t+\omega_0)+{\bf e}_2\cos(\omega t+\omega_0)\right\}.  
\end{eqnarray}
In obtaining of second term on r. h. s. we used that $[{\bf E}_{\bot}, {\bf B}]=|{\bf E}_{\bot}||{\bf B}|{\bf e}_1$, see Fig. \ref{Magn}. 

By choosing the time reference point, we can give any desired value to the constant $\omega_0$, so we put $\omega_0=0$. 
Choosing ${\bf C}_{\bot}=c[{\bf E}_{\bot}, {\bf B}]/{\bf B}^2$, the first two terms in Eq. (\ref{mf10}) vanish. This corresponds to the choice of initial position ${\bf y}(0)$ with ${\bf y}_{\bot}(0)=a{\bf e}_2$. That is at $t=0$ the particle lie in the plane $z, y$.  Then the solution reads 
\begin{eqnarray}\label{mf11}
{\bf y}_{\bot}(t)=\frac{\mu |{\bf E}_{\bot}|}{\omega}t{\bf e}_1+
a\left\{{\bf e}_1\sin\omega t+{\bf e}_2\cos\omega t\right\}\equiv {\bf O}(t)+{\bf P}(t).  
\end{eqnarray}
The vector ${\bf P}(t)$ with origin at point ${\bf O}(t)$ rotates with frequency $\omega$, while the point ${\bf O}(t)$ moves in the direction 
of ${\bf e}_1$ with the speed equal to $\mu |{\bf E}_{\bot}|/\omega$.

Choosing ${\bf C}_{\bot}=c[{\bf E}_{\bot}, {\bf B}]/{\bf B}^2-\mu|{\bf E}_{\bot}|{\bf e}_1$ and denoting  components of the vector ${\bf y}_{\bot}(t)$ in the basis ${\bf e}_1$, ${\bf e}_2$ by $x(t)$ and $y(t)$, the solution  (\ref{mf10}) reads
\begin{eqnarray}\label{mf10.1}
x(t)=\frac{\mu |{\bf E}_{\bot}|}{\omega}t+a\sin\omega t, \qquad 
y(t)=\frac{\mu |{\bf E}_{\bot}|}{\omega}+a\cos\omega t. 
\end{eqnarray}
These are parametric equations of a plane curve called trochoid, see \cite{Landau_9}. 

In resume, trajectory of a charged particle in constant and homogeneous electric ${\bf E}$ and magnetic ${\bf B}$ fields is
\begin{eqnarray}\label{mf10.2}
{\bf y}(t)={\bf y}_{\Vert}(t)+{\bf y}_{\bot}(t),
\end{eqnarray}
where the point ${\bf y}_{\Vert}(t)$ moves along the axis ${\bf B}$ according Eq. (\ref{mf5}) while the point ${\bf y}_{\bot}(t)$ moves along the  trochoid (\ref{mf11}) on the plane orthogonal to ${\bf B}$.

Let us write a variational problem for the equations (\ref{mf1}). To this aim, we introduce the scalar $A_0$ and vector ${\bf A}$ potentials at each spatial point ${\bf y}$ as follows:
\begin{eqnarray}\label{mf12}
A_0({\bf y})=({\bf y}, {\bf E}), \qquad {\bf A}({\bf y})=\frac 12 [{\bf B}, {\bf y}]=\frac 12\left(
\begin{array}{c}
B_2 y_3-B_3 y_2 \\ B_3y_1-b_1 y_3 \\ B_1 y_2-B_2 y_1 
\end{array}\right). 
\end{eqnarray}
This implies the standard agreement: ${\bf E}=-\frac{1}{c}\frac{\partial{\bf A}}{\partial t}+\nabla A_0$ and ${\bf B}=[\nabla, {\bf A}]$. Then potential  energy of the particle is
\begin{eqnarray}\label{mf13}
-U=eA_0+\frac{e}{c}({\bf A}, \dot{\bf y}), 
\end{eqnarray}
and the Lagrangian 
\begin{eqnarray}\label{mf14}
L=K-U=\frac{m}{2}\dot{\bf y}^2+eA_0+\frac{e}{c}({\bf A}, \dot{\bf y})=
\frac{m}{2}\dot{\bf y}^2+e({\bf y}, {\bf E})+\frac{e}{2c}([{\bf y}, \dot{\bf y}], {\bf B}), 
\end{eqnarray}
implies the equations (\ref{mf1}).

\end{document}